\documentstyle[11pt]{article}
\begin{document}
\thispagestyle{empty}
\begin{flushright} hep-ph/0004122\\
\end{flushright}
\vspace{0.5in}
\begin{center}

{\Large \bf Confronting Dilaton-exchange gravity with experiments }

\vspace{1.0in}

{\bf H.V. Klapdor--Kleingrothaus$^1$, H. P\"as$^1$ and U. Sarkar$^2$\\}

\vspace{0.2in}

\noindent {$^1$ \sl Max--Planck--Institut f\"ur Kernphysik,
P.O. Box 103980, D--69029 Heidelberg,
Germany \\}
\noindent{$^2$ \sl Physical Research Laboratory, Ahmedabad 380 009, India\\}

\vspace{1.0in}

\begin{abstract}

We study the experimental constraints on theories, where the 
equivalence principle is violated by dilaton-exchange contributions
to the usual graviton-exchange gravity. We point out that in this case
it is not possible to have any CPT violation and hence there
is no constraint from the CPT violating measurements in the 
$K-$system. The most stringent bound is obtained from the $K_L - K_S$ 
mass difference. In contrast, neither neutrino oscillation experiments
nor neutrinoless double beta decay imply significant constraints.

\end{abstract}
\end{center}

\newpage
\baselineskip 18pt

At present we have no indication for the the violation of 
gravitational laws. But some theories like string theory suggest
deviations from the usual graviton-exchange theories of
gravity. Thus it becomes neccessary to find out the extent of
applicability of the general theory of relativity. Several experiments
were performed to test the equivalence principle \cite{rel} for ordinary 
matter and to test local Lorentz invariance \cite{will,relc}. 
Attempts were also made to test these laws in the neutrino 
sector \cite{vep,vli,0nbb}, but these works included only  
tensorial interactions. In the K-system both tensorial and
vectorial interactions were studied by many authors \cite{vk}.

Recently it has been suggested \cite{dam} that string theory 
may lead to a different kind of
violation of the equivalence principle (VEP)
via interactions of the dilaton field, which gives an additional
contribution to the usual graviton exchange gravity. 
The resulting theory is of scalar-tensor type 
(in contrast to purely tensorial VEP discussed previously)
with the two particle
static gravitational energy
\begin{equation} 
V(r)=- G_N m_A m_B (1+\alpha _A \alpha_B)/r,
\end{equation} 
where $G_N$ is Newton's gravitational constant and $\alpha_j$ 
are the couplings of the dilaton field $\phi$ 
to the matter field of type $j$, $\psi_j$. This additional contribution may
result from a gravitational interaction 
\begin{equation} 
L= m_j \alpha_j \overline{\psi_j}\psi_j \phi.
\end{equation} 
The distinct feature of this new contribution are specific couplings
of the dilaton field to different matter fields, which violates the
equivalence principle. It has been discussed recently whether  
this feature can be tested in
 neutrino oscillation experiments 
\cite{dil,horvat}.

Unlike the violation of the equivalence principle through
tensorial interactions, in the dilaton-exchange gravity the
gravitational basis is always the same as the mass basis, since the additional
term due to dilaton exchange is directly proportional to the mass.
For this reason, a dilaton-exchange gravity cannot explain the
neutrino mixing phenomenon by itself -- the discussion in ref \cite{dil}
seems to overlook this point. However, it is
possible that a mass difference between degenerate neutrinos
is implied, what has been considered in \cite{dil}.
In this article we discuss constraints on this additional
contribution from the $K_L-K_S$ mass difference, the non-observation 
of neutrinoless double beta decay as well as   
neutrino oscillation experiments.

In a linearized classical theory one may replace the effective
mass of a fermion by
\begin{equation}
m^*_i = m_i - m_i \alpha_i \phi_c
\end{equation}
where $\phi_c$ is the classical value of the dilaton field and 
is proportional to the Newtonian potential $\phi_N$,
$$ \phi_c = \alpha_{ext} \phi_N .$$ The first term results
from the usual graviton-exchange gravity, while the second
term is the new contribution coming from the dilaton-exchange
gravity. 

To get the interaction of the usual neutrinos of different flavour,
we can rotate the effective hamiltonian in the mass and the
gravitational bases through the same unitary rotations
\begin{equation} 
H_w = U (H_m + H_g) U^{-1}  .
\end{equation}
In presence of the dilaton-exchange gravity, the effective 
hamiltonian in the mass basis is
\begin{equation} 
H_{m}+ H_g = p I + \frac{1}{2 p} {\pmatrix{
m_1 - \alpha_1 m_1 \phi_c & 0 \cr 0 & m_2
- \alpha_2 m_2 \phi_c }}^2. \label{hsew}
\end{equation}
Here $p$ denotes the momentum, $I$ represents an unit matrix, 
and for any quantity $X$ we define $\delta X = (X_1-X_2)$ and 
$\bar{X} = (X_1+X_2)/2$.

Assuming there is no $CP$ violation, the
effective hamiltonian becomes real and symmetric and we can
parametrize the mixing matrix by
\begin{equation}
U = \pmatrix{ \cos \theta & \sin \theta \cr -\sin \theta & \cos 
\theta}
\end{equation}
where $\theta$ is the mixing angle. Then the effective hamiltonian
in the weak basis becomes
\begin{equation}
H_w = p I + {1 \over 2 p} {\pmatrix{M_+ & M_{12} 
\cr M_{12} & M_-}}^2
\end{equation}
with
\begin{eqnarray}
M_\pm &=& \bar{m} + \frac{1}{2}(\alpha_1 m_1 + \alpha_2 m_2) \phi_c \pm
\frac{1}{2}[\delta m + (\alpha_1 m_1 - \alpha_2 m_2) \phi_c] \cos 2 \theta
\nonumber \\
M_{12} &=&  -\frac{1}{2}[ \delta m + (\alpha_1 m_1 - \alpha_2 m_2) \phi_c] 
\cos 2 \theta \label{pm}  .
\end{eqnarray}
The mass squared difference between the physical masses is
given by
\begin{equation}
\Delta m^{*2} = \Delta m^2 - 2 \phi_c ( \alpha_2 m_2^2 - 
\alpha_1 m_1^2) + \phi_c^2 (\alpha_2^2 m_2^2 - 
\alpha_1^2 m_1^2) ,  \label{ms}
\end{equation}
while the mass difference is given by
\begin{equation}
m^{*}_2 - m^{*}_1 = \delta m + (\alpha_2 m_2 - 
\alpha_1 m_1) \phi_c  \label{md} .
\end{equation}

We shall now use the above formalism to analyse the constraints
on dilaton-induced gravity from the $K_L-K_S$ mass
difference, neutrinoless double beta decay and neutrino 
oscillation experiments. For the $K-$system, the physical states are
the $K_L$ and $K_S$ states, while the weak states are the
$K^\circ$ and $\overline{K^\circ}$ states. Thus the $K_L-K_S$
mass difference can be read off from equation [\ref{md}],
\begin{equation}
m^*_L - m^*_S = (m_L - m_S) - m_L \phi_c ( \alpha_L  
- {m_S \over m_L} \alpha_S ) .
\end{equation}
The experimental value $m^*_L - m^*_S$
is dominated by the mass difference $m_L-m_S$. Thus
no significant cancellations are expected and a conservative bound for
the contribution from dilaton exchange is
$$ |m_L \phi_c ( \alpha_L  
- {m_S \over m_L} \alpha_S )| < m^*_L - m^*_S.$$
Using also $m_S \simeq m_L$ this implies
a bound on the dilaton-induced 
gravity coupling of 
\begin{equation}
\delta \alpha < {1 \over \phi_c} \left( {m_L - m_S \over m_L}\right) _{expt}
= {7 \times 10^{-15} \over \phi_c},
\end{equation}
where we used ${m_L - m_S \over m_L} \sim
7 \times 10^{-15}$ \cite{pdg}.
While the bound depends on the absolute value of $\phi_c$, we
can obtain a rough estimate by considering the value of the Newtonian
gravitational potential to be due to the great attractor, which is
about $3 \times 10^{-5}$ and $\alpha_{ext} \sim 0.03$, so that
$\phi_c \sim 10^{-6}$. Then the bound becomes  
$\delta \alpha < 7 \times 10^{-9}$. 

Unlike for the tensorial interactions, in this case the measurement of 
the CPT violating parameter, the mass difference of 
$K^\circ$ and $\overline{K^\circ}$, does not yield any constraint.
This can be understood by considering 
the mass difference,
which can be read off from equation (\ref{pm}) to be
\begin{equation}
{M_+ - M_-} =  \delta m \cos 2 \theta
+  \phi_c \cos 2 \theta (\alpha_2 m_2 - \alpha_1 m_1)  .
\end{equation}
Here the first contribution is the usual mass contribution and the
second contribution is the one coming from dilaton exchange.
Since there is no CPT violation in the usual gravity, we have
$\cos 2 \theta =0$. This implies that there is no 
contribution coming from the dilaton exchange gravity. 

We now turn over to discuss the constraints coming from the
neutrino sector. The decay rate for neutrinoless double beta
decay is given by,
\begin{equation} 
[T_{1/2}^{0\nu\beta\beta}]^{-1}=\frac{M_+^2}{m_e^2} 
G_{01} |ME|^2,
\end{equation} 
where $ME$ denotes the nuclear matrix element $ME=M_F-M_{GT}$, 
(for numerical values see \cite{mat}), $G_{01}$
corresponds to the phase space factor defined in \cite{doi}
and $m_e$ is the electron mass. 
In contrast to the tensorial VEP \cite{0nbb} for dilaton exchange gravity the
observable has no explicit momentum dependence.
The contribution of the dilaton exchange to the observable for 
neutrinoless double beta decay is given by
\begin{equation}
M_{+ dil} = \frac{1}{2}m_2 \phi_c [({m_1 \over m_2} \alpha_1 + \alpha_2)
+ ({m_1 \over m_2} \alpha_1 - \alpha_2) \cos 2 \theta] .
\end{equation}
Since no momenta enter the observable, the decay rate is suppressed 
considerably (see also the discussion in \cite{richtig}, which critizizes
the treatment in \cite{wrong}).

The dominant contribution to the neutrino oscillations comes 
through the change in the mass squared difference, given by
\begin{equation}
<\Delta m^2>_{dil} = -2 \phi_c m_2^2 (\alpha_2 -
{m_2^2 \over m_1^2} \alpha_1) .
\end{equation}
In the almost degenerate case the bounds from the two experiments
can be compared if we assume that there is no mean deviation of usual
gravity $(\alpha_1 + \alpha_2) = 0$, which is assumed in most cases. 
An alternative natural choice is to assume 
that the masses of the neutrinos are hierarchical to explain the
atmospheric and solar neutrino problems. In the case of hierarchical
neutrino masses, $m_2 \gg m_1$, $\alpha_1$ drops
out from both expressions and without any further 
assumption all experiments can be compared in terms of the
single unknown parameter $\alpha_2$. 
In this case it is even more difficult to obtain any bound from neutrino
experiments since for neutrino masses $m_i$ only upper bounds exist. 

If we consider according to ref. \cite{dil}
the upper bound of 
$\alpha_{ext} \sim \sqrt{10^{-3}}$ as its absolute value, 
and the dominant 
contribution to the local gravitational potential to be due to the
great attractor, which is about $\phi_N=3 \cdot 10^{-5}$,
we can estimate the bounds on the dilaton couplings.
Although this bound cannot be taken seriously, this allows 
us to compare our result with earlier results.
Only in the almost degenerate case ($m_1 \sim m_2 \sim m$), 
assuming $m=2.5$ eV (as an upper  bound obtained 
from tritium beta decay experiments \cite{tritium})
and only for the case the vacuum oscillation solution of the solar neutrino
problem will turn out to be realized in nature,
the experimental $\Delta m^{*2}$ may be large enough to imply a significant 
bound (see however the discussion of medium effects in \cite{horvat}). 
The search for neutrinoless double beta decay 
\cite{double},
implying $M_{+ dil} 
<0.3$ eV suffers from an even more severe suppression, since the bound for 
$M_+$ is less stringent than the one for $\Delta m^{*2}$.

In summary, we point out that the dilaton exchange gravity cannot
be constrained substantially in the neutrino sector, while the bound 
coming from the $K_L - K_S$ mass difference is significant, 
modulo the uncertainty of classical background potential. Unlike the 
new tensorial or vectorial gravitational interactions, this scalar 
interaction cannot introduce any CPT violation in the $K-$system.

\end{document}